\documentclass[12pt,preprint]{aastex}
\usepackage{amsmath}
\def\dfplot#1{\plotone{#1}}
\def\BE{\begin{equation}}
\def\BEL#1{\begin{equation}\label{#1}}
\def\EE{\end{equation}}
\newcommand{\IRAS}{{\it IRAS}}

\newcommand{\WMAP}{{\it WMAP}}

\newcommand{\HII}{H\,{\scriptsize II}}

\newcommand{\Halpha}{H$\alpha$}
\newcommand{\etal}{{\it et al.}~}

\newcommand{\degree}{^\circ}
\newcommand{\s}{{\rm ~s}}

\newcommand{\g}{{\rm ~g}}
\newcommand{\cm}{{\rm ~cm}}
\newcommand{\km}{{\rm ~km}}

\newcommand{\GHz}{{\rm ~GHz}}

\newcommand{\K}{{\rm ~K}}

\newcommand{\microK}{\mu{\rm K}}
\newcommand{\eV}{{\rm ~eV}}
\newcommand{\GeV}{{\rm ~GeV}}
\newcommand{\pc}{{\rm ~pc}}
\newcommand{\kpc}{{\rm ~kpc}}
\newcommand{\Mpc}{{\rm ~Mpc}}

\newcommand{\vecx}{{\bf x}}
\newcommand{\matM}{{\bf M}}

\newcommand{\erg}{{\rm ~erg}}
\newcommand{\sigmav}{\langle\sigma_Av\rangle}

\begin{document}
\title{WMAP Microwave Emission Interpreted as
Dark Matter Annihilation in the Inner Galaxy}

\author{Douglas P. Finkbeiner\footnote{Henry Norris
Russell Fellow, Cotsen Fellow}}
\affil{Princeton University, Department of Astrophysics,
Peyton Hall, Princeton, NJ 08544}

\begin{abstract}
Excess microwave emission observed in the inner Galaxy (inner $\sim 1-2$
kpc) is consistent
with synchrotron emission from highly relativistic $e^+e^-$ pairs
produced by dark matter particle annihilation.  More conventional
sources for this emission, such as free-free (thermal bremsstrahlung),
thermal dust, spinning dust, and the softer Galactic synchrotron
traced by low-frequency surveys, have been ruled out.  The total power
observed in the range $23 < \nu < 61\GHz$ is between $1\times 10^{36}$
and $5\times 10^{36}\erg\s^{-1}$, depending on the method of
extrapolation to the Galactic center, where bright foreground emission
obscures the signal.  The inferred electron energy distribution is
diffusion hardened, and is in qualitative agreement with the energy
distribution required to explain the gamma ray excess in the inner
Galaxy at $1-30\GeV$ as inverse-Compton scattered starlight.  We
investigate the possibility that this population of electrons is
produced by dark matter annihilation of $100\GeV$ particles,
with cross section $\sigmav=2\times10^{-26}\cm^3\s^{-1}$, and an
$r^{-1}$ dark matter mass profile truncated in the inner Galaxy, and
find this scenario to be consistent with current data.
\end{abstract}


\section{INTRODUCTION}

It is almost universally accepted that the majority of matter in the
Universe is non-baryonic.  Existence of this ``dark matter'' is
supported by several kinds of evidence: galaxy rotation curves,
gravitational lensing of background objects by galaxy clusters, the
small value of $\Omega_b h^2$ determined from big bang
nucleosynthesis, and $\Omega_b/\Omega_m$ from the cosmic microwave
background (CMB) anisotropy.  The gravitational effects of dark matter
are observed, but in spite of numerous investigations with particle
accelerators and direct detection experiments, no direct signature of
dark matter interaction with ordinary matter has ever been found.

Supersymmetry theory (SUSY) provides a candidate dark matter particle:
a linear combination of higgsino, Z-ino, and photino states commonly
known as the neutralino, $\chi$, the lightest stable supersymmetric
particle (See Jungman, Kamionkowski, \& Griest 1996 for a
review).  Direct detection of neutralinos is difficult because of
their very weak interactions with ordinary matter, but if the
neutralino is a Majorana fermion (and therefore is its own
anti-particle, $\chi=\bar{\chi}$), it self-annihilates and produces
some combination of W and Z bosons, mesons, $e^+e^-$ pairs, and
$\gamma$-rays (see Gunn \etal\ 1978 for the basic scenario).  Baltz \&
Wai (2004) point out
that a significant fraction of the power liberated in
self-annihilation may go into ultra-relativistic $e^+e^-$ pairs
(hereafter ``electrons''), producing observable microwave synchrotron
radiation.  These high energy electrons inverse Compton scatter
CMB and starlight photons, shifting them to soft and hard
$\gamma$-rays, respectively.  To date, efforts have focused largely on
$\gamma$-ray observations of dark matter concentrations in dwarf
spheroidal galaxies (Gondolo 1994, Blasi \etal\ 2003, Baltz \& Wai
2004) or the
Galactic center (Stoehr \etal\ 2003), or synchrotron observations of
satellite galaxies (Baltz \& Wai 2004).  Direct detection of high
energy positrons near Earth has also been discussed (Baltz \& Edsj\"o 1998,
Baltz \etal\ 2002), but is thought to be unlikely, given the
foreground produced by other mechanisms and the high boost factors
required to obtain an observable signal (Hooper \etal\ 2004).
Copious annihilation near the black hole in
the Galactic center would be detectable at radio frequencies for most
neutralino models if there
were a steep central cusp in the density profile (Gondolo 2000)
but non-detection may indicate lack of a cusp.
Detection of synchrotron emission near the black hole in
the Galactic center has also been considered (Bertone \etal\ 2002, Aloisio
\etal\ 2004) but detection of the synchrotron annihilation signal in
the inner regions ($\sim 1$ kpc) of the Galaxy was thought to be
impractical because of the complexity of the ISM microwave emission
from other components (thermal dust, spinning dust, free-free, and
``ordinary'' synchrotron from shock-accelerated electrons).  Recent
advances in our understanding of ISM emission have made this problem
tractable.

The \emph{Wilkinson Microwave Anisotropy Probe} (\WMAP) has revealed
the structure of the Galactic ISM with unprecedented precision at
microwave ($22-94\GHz$) frequencies, leading to a reassessment of the
Galactic foregrounds (Bennett \etal\ 2004, Finkbeiner 2004, Finkbeiner
\etal\ 2004).  One
surprising result of the increasingly careful foreground analysis is
an excess of emission in the inner Galaxy, distributed with
(approximate) radial symmetry within $\sim 20\degree$ of the Galactic center
(dubbed the ``haze,'' Finkbeiner 2004).
This emission is not correlated with any of the foreground templates
representing the known microwave foregrounds.  Although the spectrum of
this excess is somewhat flatter than standard synchrotron emission, we
argue that it is consistent with a population of ultra-relativistic
electrons created in the inner Galaxy and diffusing outward, and
consider the possibility that these electrons are products of
dark matter (DM) annihilation.  To be concrete, we refer to the
DM particle as the neutralino throughout, although \emph{any} stable
neutral particle with a mass of $\sim 100\GeV$ and the assumed
$\sigmav$ would have the consequences described below.

Assuming the neutralino is the dominant form of dark matter, 
the average relic density today is $\Omega_\chi h^2 =
(\Omega_m-\Omega_b) h^2 = 0.113\pm.009$ according to the \WMAP\
cosmology (Spergel \etal\ 2003).
By setting the annihilation rate equal to the expansion rate after
freeze-out, the cross section\footnote{
The parameter $\sigmav$ is commonly refered to as the ``cross section''
even though that term would properly refer only to $\sigma_A$.  It
would be more precise to call $\sigmav$ the ``annihilation rate
coefficient.''}  $\sigmav$ is determined by this
relic abundance (see \S \ref{sec_relic}).  Using a fiducial mass of
$100\GeV$, a diffusion term ($K$) derived from cosmic ray data, and an
energy loss term ($A E^2$) including synchrotron losses and inverse Compton
scattering (ICS) off of starlight and CMB photons, we calculate the
electron energy distribution as a function of Galactocentric radius
for various assumptions.  An analytic solution is presented for the
simplest case, and numerical solutions for variable $A$ and $K$
terms. 

For the assumed dark matter distribution, all models produce a similar
(factor of $\sim 2$)
power output, since the power input by
annihilation is fixed \emph{a priori}.  In all cases, the spectrum,
morphology, and predicted synchrotron power output by the model is
similar to the observed \WMAP\ excess.  This is tantalizing, because
if one can determine the losses (i.e. energy density of starlight and
magnetic field) in the inner Galaxy, the fraction of power liberated
as relativistic $e^+e^-$,
and the dark matter density profile, then the total $\chi$
annihilation power can be computed.  Given the dark matter
distribution and the annihilation power in the inner Galaxy, the
approximate mass, $m\chi$, can be computed.  With luck, an
annihilation line or sharp cut off in $\gamma$ ray continuum will be
observed, fixing the mass more precisely.  With refined Galactic
models and improved microwave and $\gamma$ ray data, it may actually
be possible to determine the mass and cross section of a new particle
astrophysically, making a specific prediction to be verified by
accelerator experiments (e.g. at the LHC, Ellis \etal\ 2004).

The main difference between our model and previous models is that
electron diffusion is solved for in a self-consistent way (like Baltz
\& Wai 2004) and its effect on the synchrotron spectrum computed.  The
consequent flattening of the spectrum, especially near the Galactic
Center, has not been emphasized in the previous literature. 

In \S 2, the excess microwave emission near the Galactic Center is
described.  Model parameters for the hot electron population and ISM
are justified in \S 3, and the model is evaluated and compared to
\WMAP\ data in \S 4.

\section{THE \WMAP\ HAZE}

The \emph{Wilkinson Microwave Anisotropy Probe} (\WMAP ) has observed
the sky with good sensitivity ($\sim150\microK$ per pixel in the 1 yr data)
and measured the cosmological anisotropy power spectrum (Hinshaw
\etal\ 2003) and, with less significance, its polarization (Page
\etal\ 2003), in order to determine fundamental cosmological
parameters (Spergel \etal\ 2003).  To the extent that Galactic
foreground signals from thermal dust emission, free-free (thermal
bremsstrahlung), ``ordinary'' synchrotron (relativistic electrons
accelerated by supernovae and spiraling in the Galactic magnetic field) and
spinning dust (see Draine \& Lazarian 1998) are present in the data,
they have been removed and do not interfere with the cosmological
signal (Bennett \etal\ 2003).

In the course of studying these foregrounds and exploring the
possibility of spinning dust (Finkbeiner 2004, Finkbeiner \etal\ 2004),
evidence for an additional emission component has
emerged.  This enigmatic emission is centered at the Galactic center
and extends $20-30\degree$ away in all directions, falling rapidly
($\sim 1/R$) with projected distance from the Galactic center.  It is
not likely to be an artifact of the foreground subtraction, because
south of the GC it is similar in brightness to the other components,
and the fit is sufficiently rigid (only eight parameters for
the whole sky to fit 5 bands $\times3\times10^6$ pixels) that it is
robust to errors in the templates.  Even north of the GC, the excess
is approximately symmetrical about the GC in spite of much brighter
foreground objects such as the large ($\sim 10\degree$) nearby nebula
of $\zeta$Oph.  This emission component has
been nicknamed ``the haze'' to avoid giving it an interpretive name
prematurely (see Finkbeiner 2004).

The haze was first thought to be a free-free component from gas too
hot to have recombination line emission ($T >> 10^4\K$) and too cold
to have significant X-ray emission in the ROSAT (Snowden \etal\ 1997)
1.5 keV band  ($T << 10^6\K$).  North of the GC there is sufficient
gas and dust to hide
the \Halpha\ recombination line and X-ray emission, but to the South,
column densities are low enough that both would be seen.  For
$\sim10^5 \K$ gas the X-ray emission would be soft enough that gas
absorption could block it, but gas is thermally unstable at $10^5$ K
because of efficient metal-line cooling
(e.g., Spitzer 1978), and the amount of gas required is inconsistent with
\HII\ column densities determined from pulsar dispersion measures
given by Taylor \& Cordes (1993).  Furthermore, the free-free spectrum
of $10^5$ K \HII\ extends up to $\nu \sim 10^7\GHz$ and the 
integrated spectral power suggests a substantial free-free power output of
$1-5\times 10^{40}\erg\sec^{-1}$.  It
is necessary to consider alternative sources of emission.


\section{PARAMETERS}
In this section, the model parameters are specified: annihilation
cross section, mass, $e^+e^-$ energy distribution, magnetic field
strength, photon field available for ICS, and dark matter mass profile
in the Milky Way.  The parameters are listed in Table 1. 

\subsection{Neutralino Cross Section}
\label{sec_relic}

If the neutralinos present today are the cosmological relic of thermal
production in the early Universe, then the mean neutralino mass
density as a fraction of the critical density, $\Omega_\chi$ can be
computed from its mass and cross section.\footnote{
If there is more than one WIMP particle with
significant abundance at freeze-out, one must then consider
co-annihilations and
other interactions, and the situation is considerably more complex
(e.g. Griest \& Seckel 1991, Edsj\"o \& Gondolo 1997, Ellis \etal\
2000).}  The \WMAP\
cosmology (Spergel \etal\ 2003) gives $\Omega_{DM} h^2 =
(\Omega_m-\Omega_b) h^2 = 0.113\pm.009$, where the subscripts ``DM'', ``m''
and ``b'' denote ``dark matter'', ``total matter'' and ``baryon'',
respectively, and $h$ is the
Hubble constant in units of $100\km \s^{-1}\Mpc^{-1}$. 

The neutralino number density falls rapidly when the
temperature of the Universe ($T$) falls much below the particle mass
($m_\chi$).  After this, neutralino annihilations cause the
number density, $n$, to drop rapidly [$n \sim \exp(-m_\chi c^2/kT)$] to
the level where the particle
annihilation rate ($\Gamma$) is comparable to the expansion rate,
$H(t)\sim\dot{a}/a \sim t^{-1}$, after which the comoving density
asymptotes to the ``freeze-out'' density.
In the early, radiation-dominated,
Universe, $t\sim a^2$, so the criterion is
\BE
\Gamma = \sigmav n \sim \sigmav n_0 a^{-3} \sim H \sim a^{-2}
\EE
where $n_0=na^3$ is the co-moving number density.  Because temperature
$T\sim a^{-1}$, 
\BE
n_0 T_f \sim \frac{1}{\sigmav}
\EE
where $T_f\sim m_\chi$ is the freeze-out temperature, proportional to
the particle mass.  This leads to the remarkable conclusion that the
co-moving relic mass density, $\rho_\chi=n_0 m_\chi$, is inversely
proportional to the cross section, so if one knows the relic mass
density today (or assumes it) the cross section is fixed.  A more
complete calculation (Jungman \etal\ 1996) yields the same relation in
terms of $\rho_\chi\sim\Omega_\chi h^2$:
\BEL{eq_relic}
\Omega_\chi h^2 = \frac{3\times 10^{-27}\cm^3\s^{-1}}{\sigmav}
\EE
A proper calculation of the relic density takes account of the form of
the density drop after freeze-out and introduces a weak mass
dependence in the relic density, but the approximation given in
Eq. (\ref{eq_relic}) is adequate for our purposes, and we adopt
$\sigmav = 2\times 10^{-26}\cm^3\s^{-1}$.

\subsection{Neutralino Mass}
We take the neutralino mass, $m_\chi$ to be $100\GeV$, approximately
at the weak scale, and lower than the bounds of $m_\chi < 600\GeV$
given by Ellis \etal\ (2000), or $m_\chi < $ several TeV (Baltz \&
Gondolow 2004).
This choice of mass is arbitrary, and is
degenerate with the fraction of power going into energetic $e^+e^-$
given in \S\ref{sec_result}.

\subsection{Energy Distribution}
There is considerable uncertainty in both the injection energy
distribution and the effects of diffusion on the resultant number
density per energy, $n(E,r)$.  One possible assumption is that
annihilating neutralinos go directly to quark anti-quark pairs which
then go to pions:
$$
\chi\bar{\chi} \rightarrow \gamma + \pi^o, \pi^+, \pi^-
$$
The pions then decay:
$$
\pi^+ \rightarrow \mu^+ + \nu_\mu
$$
$$
\pi^- \rightarrow \mu^- + \bar{\nu}_\mu
$$
followed by
$$
\mu^+ \rightarrow e^+ + \nu_e + \bar{\nu}_\mu
$$
$$
\mu^- \rightarrow e^- + \bar{\nu}_e + \nu_\mu
$$

This results in an injection energy distribution that goes as $n_i\sim
E^{-3/2}$ in the low energy limit (see Blasi \etal\ 2003), with a
significant fraction of the energy carried away by neutrinos.  
However, if annihilation to $\tau$ leptons is important
($\chi\bar{\chi} \rightarrow \tau^+\tau^-$)
the resulting $e^+e^-$ spectrum would be harder (Hooper \etal\ 2004).

The injection spectrum must then be evolved through the diffusion and
energy loss equations to determine the steady state electron density. 
The model in \S\ref{sec_model} assumes spherical symmetry, but one
might also consider 
cylindrical symmetry with boundaries a few kpc from the Galactic
midplane at which the electrons escape.  If the electrons have time to
diffuse to the boundaries before losing most of their energy, the
electron spectrum can be considerably hardened.  In the limit of rapid
escape, the steady state spectrum is just the injection spectrum, so
that sets a limit on the spectral hardness. 

\subsection{Photon energy density}
\label{sec_photons}
The starlight photon energy density in the inner Galaxy is necessary to compute
electron energy loss due to inverse Compton scattering.  
Strong \etal (2000) have considered a model based on \IRAS\ and DIRBE
infra-red data in addition to the usual stellar information, and find
that the starlight energy density is approximately
$$
U_{*} = 11 e^{-r/3\kpc} \eV\cm^{-3}
$$
in the Galactic plane, or about $0.6\eV\cm^{-3}$ at $r_\sun$.
This compares to the energy density of $0.6\eV\cm^{-3}$ for a
$\bar{B}=5\mu$G magnetic field, 
and $0.260\pm0.001\eV\cm^{-3}$ for the CMB.
The far IR emission is subdominant, but should be included in future
models.

\subsection{The Galactic Magnetic Field}
Heiles (1996) presents measurements of different moments of the
Galactic magnetic field in the plane at Galacto-centric distances of 4
and 8.5 kpc.  None of these measurements is exactly the quantity we
need, $\bar{B}=\langle B^2 \rangle^{-1/2}$, but they are consistent
with $\bar{B}=5\mu$G at the Solar circle and increasing exponentially
toward the
Galactic center.  To demonstrate that our result is not
strongly field dependent, we try two cases: $\bar{B}=5\mu$G, and
$\bar{B}=21 e^{-r/6\kpc} \mu$G.  This field is similar to the
field radial dependence derived by Han (priv. comm.) with a scale
length of $7.5\kpc$.  Because the $B$ field and starlight are so close
to equipartition, we assume that they are exactly in equipartition for
simplicity, even though there is no obvious mechanism to regulate
this.  Also, the field is assumed to be tangled in an isotropic way,
so that the particles move in a random walk.

A more detailed model would incorporate what is known about field
strength variation with position in the Galaxy, especially the fact
that the field is stronger in the spiral arms than in inter-arm
regions.  Synchrotron emission is enhanced perpendicular to the field
lines and suppressed along them; we assume isotropic tangling of the
field and therefore isotropic emission.
 Furthermore, the ratio of the tangled to ordered strength,
and the mean direction of the ordered part may be important.  For
example, Han (2004) suggests that the bulge is dominated by a
solenoidal component perpendicular to the Galactic plane, which may
facilitate rapid escape of electrons from the Galactic center, whereas
particles created in the arms may be trapped there for longer periods
of time.  Therefore we should be cautious about applying conclusions
drawn from local cosmic ray propagation to particles in the inner
Galaxy.  This is a rich subject, and is well beyond the scope of this
article.

\subsection{Mass profile}

We assume a truncated Navarro, Frenk, \& White (1997; NFW) profile.
Because we are only interested in the inner $1-2\kpc$ of the Galaxy,
the larger scale shape of the NFW profile is unimportant to this
calculation, other than that it serves to tie the assumed Milky Way
mass (within 400 kpc) of $2\times10^{12} M_\sun$ to a mass density
at the center.  We assume an NFW profile with $r_s=20$ kpc, and
arbitrarily truncate the core at 600 pc.  
Although recent computer simulations indicate
that NFW is not valid at the core \cite{stoehr03}, there is
theoretical support from the Jeans equation for an index of -1 or
steeper in the core (Hansen 2004).  In the case of a Moore (1998)
profile ($\rho\sim r^{-1.4}$) the annihilation power can be a factor of 5 times
higher. 
The details of the central truncation 
are unimportant for NFW profile (where each concentric shell
contributes equal annihilation power) but becomes important for the
Moore profile. 


\section{THE MODEL}
\label{sec_model}


\subsection{Analytic Solution}
\label{sec_analytic}
At energies of tens of GeV, electrons moving in $\mu$Gauss fields take
millions of years to lose half their energy, so it is a poor
approximation to neglect electron diffusion.  We assume that electrons
resulting from the annihilation are created at a rate proportional to
$\chi$ number density squared and then diffuse in a random walk
through the (rather tangled) Galactic magnetic field.  The number density
$n(E,\vecx) dE$ between energy $E$ and $E+dE$ evolves according to 
\BEL{eq_pde}
\frac{d}{dt} n(E, \vecx) = \nabla \cdot (K(E,\vecx)\nabla n)
+\frac{\partial}{\partial E}\left[b(E,\vecx) n\right]+Q(E,\vecx)
\EE
where $K(E,\vecx)$ is the diffusion coefficient (area per time),
$b(E,\vecx)$ is the energy loss coefficient (energy loss per time),
and $Q(E,\vecx)$ is the source term (number density per time). 
Neglecting the position dependence of $K$ and $b$, the steady state
equation is:
\BE
-K(E) \nabla^2 n
-\frac{\partial}{\partial E}\left[b(E) n\right] = Q(E,\vecx).
\EE
In the simple case where $K$ is energy independent and $b=-AE^2$
with $A$ constant, the
solution for a mono-energetic delta-function source at the origin,
$Q(E,\vecx) \sim \delta(\vecx)\delta(E-E_0)$, and zero density at
$r=\infty$, is simply a superposition of spherically symmetric
Gaussians losing energy and
expanding with $\sigma^2 \sim t$.  The number
density of electrons produced between $t$ and $t+dt$ in the past is
\BE
n(E(t),r) dE/dt =
  Q_0\frac{1}{[2\pi\sigma(t)^2]^{3/2}}\exp(-r^2/2\sigma(t)^2)
\EE
with $Q_0$ the creation rate (number per second),
and where electrons created with energy $E_0$ at $t=0$ lose energy as
\BE
b(t)=\frac{dE}{dt}=-AE^2
\EE
so that
\BE
E=\frac{E_0}{(1+AE_0t)}.
\EE
In general, the energy loss rate of each electron is
\BE
b=\frac{4}{3}\sigma_Tc\gamma^2\beta^2(U_B+U_{\rm ph})
\EE
where $\sigma_T$ is the Thomson cross section, $\beta=v/c$ is the
speed of the electron, $\gamma mc^2$ is the
electron energy, $U_B=B^2/8\pi$ is the magnetic field energy density
and $U_{\rm ph}$ is the photon energy density.  In the ultra-relativistic
limit,
\BEL{eq_A}
A=\frac{4\sigma_T(U_B+U_{\rm ph})}{3m^2c^3},
\EE
where $U_{\rm ph}=U_{\rm star}+U_{\rm CMB}$.
For electrons created time $t$ ago, 
\BE
\sigma^2=2kt=\frac{2k(E_0-E)}{AE_0E}
\EE
so that the number density as an explicit function of energy is:
\BEL{eq_green}
n(E,r) = Q_0 \left[\frac{AE_0E}{4\pi k(E_0-E)}\right]^{3/2}
\frac{1}{AE^2}\exp\left(\frac{-AE_0Er^2}{2k(E_0-E)}\right)
\EE
and the integral along the line of sight is
\BE
N(E,R) = 
\frac{Q_0E_0}{4\pi k E (E_0-E)}
\exp\left(\frac{-AE_0ER^2}{2k(E_0-E)}\right)
\EE
where $R$ is the projected Galacto-centric distance.
This Green function solution must be convolved with the projected
source density to obtain the solution for arbitrary $Q$, but just this
simple expression provides some intuition.  (See Appendix A for more
details).
The energy distribution of
the entire $e^-$ population is $n(E)\sim E^{-2}$, as one would obtain
for a population created at energy $E_0$ and confined to a box.
However, along any line of sight, the distribution is flatter ($\sim
E^{-1}$) because of diffusion.  The older electrons that have lost
much of their energy have also had time to diffuse away from the
source, leaving a flatter spectrum below some cut-off energy and a
sharp drop above it.
This cut-off energy decreases with distance from the
source.  Another feature of the above solution is that the total
projected radial
density $N(R) = \int dE~n(E,R) \sim r^{-1}$ diverges in the absence of
a lower energy bound.  This makes sense, because electrons have been
created at rate $Q_0$ forever and never disappear.  In
reality, the diffusion zone in the Galaxy has finite size, and most
electrons escape while still relativistic.  The effects of this escape
will be seen in the numerical solutions in the next section.

\subsection{Numerical Solutions}
\label{sec_numerical}
For less restrictive assumptions about $K$, $A$, and $Q$, analytic
solutions to Eq
(\ref{eq_pde}) are cumbersome or non-existent, and the PDE must be
solved numerically.  Being interested in the inner Galaxy, where
dark matter density is assumed to be highest, we assume
spherical symmetry and use the fact that the Laplacian of a radial
function can be written
$$
\nabla^2n(r)=\frac{1}{r}\frac{\partial^2}{\partial r^2}(rn).
$$
It is convenient to solve for $r n(E,r)$ on a radius-energy $(r,E)$ grid
and divide the result by $r$.  In most cases a grid with 30
logarithmic energy bins $(0.1 < E < 100\GeV)$ and 60 radial bins
$(0 < r < r_{\rm max} = 12000\pc)$ works well.  Boundary conditions
for $r n$ are
taken to be zero at $r=0$ (as required) and $r=r_{\rm max}$ and an NFW
profile squared as the input source distribution at maximum energy.
The density is truncated in the inner 600 pc to the value at 600 pc to
avoid the central cusp.  Results are not
sensitive to this truncation. 

The matrix $\matM$ describing the evolution of the
system in state $S$ (i.e. $r$ times density at discretely sampled
positions and energies) from time step $i$ to $i+1$ is constructed such
that
$$
S_{i+1}=S_i + \matM S_i.
$$
The steady state solution is the nullspace of $\matM$ (i.e. $\matM S=0$),
readily found with SVD (singular-value decomposition).  This is
in general a multi-dimensional space providing exactly enough freedom
to fit the boundary conditions stated above.

Grid solutions have been computed for two cases (in each case,
$K_0=3\times10^{-27}\cm^2 \s^{-1}$ and $A_0=10^{-16}\s^{-1}\GeV^{-1}$):
(1) $K=3K_0, A=A_0$, and
(2) $K(E) = 3K_0(3^\alpha + (E/1\GeV)^\alpha$), $\alpha=0.6$, and
$A$ representing the energy loss from synchrotron and scattering of
CMB and starlight photons assuming equipartition of starlight and the
magnetic field, and an exponential
radial dependence (see Eq. \ref{eq_A}).

The energy dependence of $K(E)$ given here is proportional to but
somewhat higher than that found in the study of
cosmic ray propagation in the Galaxy, including diffusion and
convection, by Webber, Lee, \& Gupta (1992).  Although this study
considered cosmic-ray nuclei and not leptons, at the high particle
rigidity (momentum per charge) of interest, $e^+e^-$ diffuse in a
similar fashion.

%

\begin{figure}[tb]
\epsscale{0.8}
\dfplot{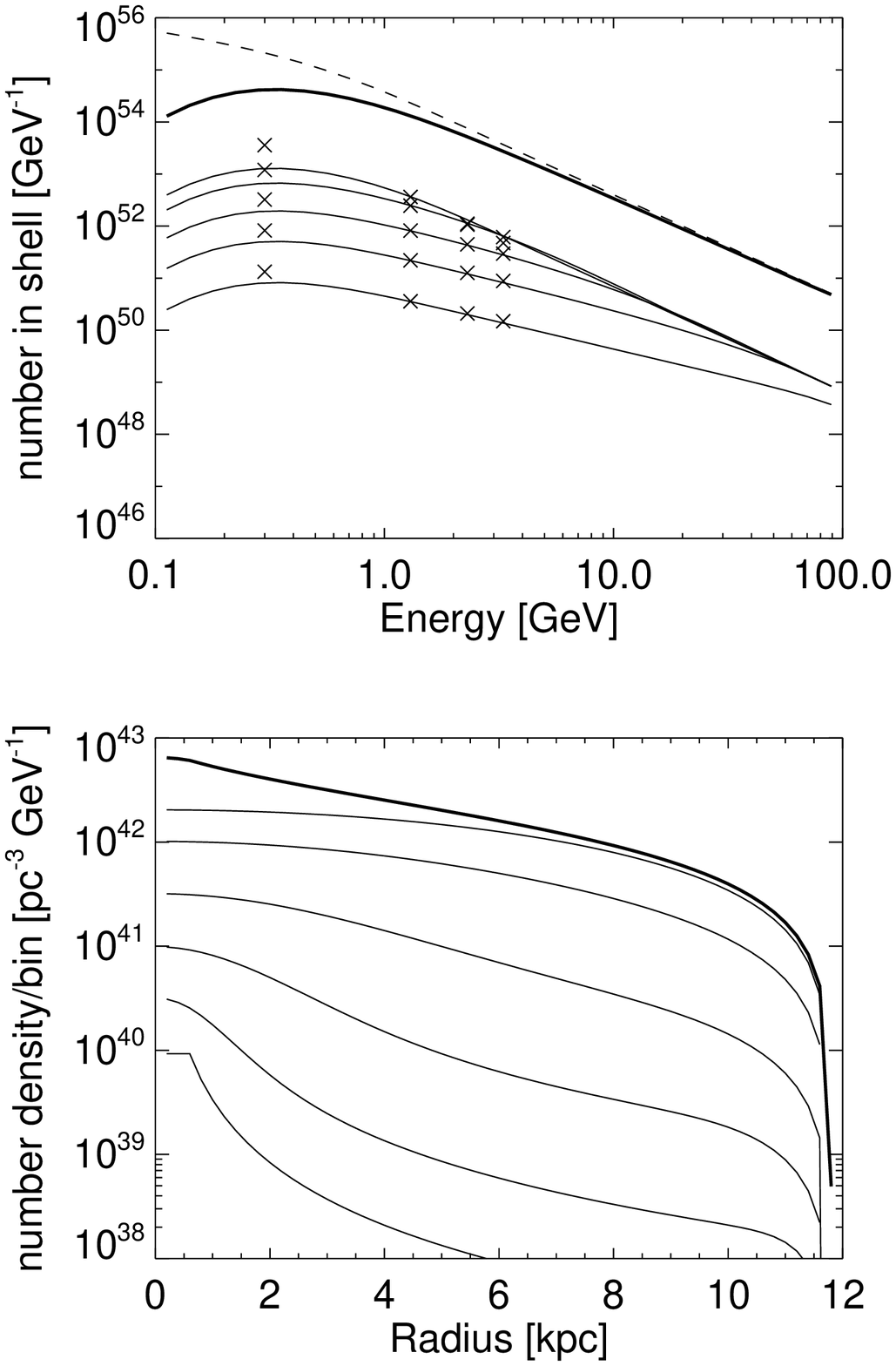}
\figcaption{(\emph{top}) Model 1 electron energy distribution for
all electrons (\emph{thick line}) and spherical shells of thickness
$\Delta r=200\pc$ at radii of 8, 4, 2, 1, and 0.4 kpc (\emph{thin
 lines, top to bottom}) from the numerical calculation in
\S\ref{sec_numerical}.  For comparison, the analytic solution in
\S\ref{sec_analytic} is evaluated for all electrons (\emph{dashed
 line}) and shells at the same radii (\emph{crosses}).
The numerical solution is a good approximation to the analytic
solution at $E>1\GeV$ and inner radii, where the \WMAP\ emission is
evaluated. 
(\emph{bottom}) Number density ($\pc^{-3}$) as a function of
Galacto-centric radius for all electrons (\emph{thick line}) and for
logarithmically spaced bins of energy 0.35, 1.1, 3.5, 11, 35, and 89
GeV (\emph{thin lines, top to bottom}). 
The highest energy bin is truncated at 600 pc.  In both panels the
figures are normalized assuming a total annihilation rate of $4\times
10^{38}\s^{-1}$ as found in Appendix A.
\label{fig_simple}
}
\end{figure}

\begin{figure}[tb]
\epsscale{0.8}
\dfplot{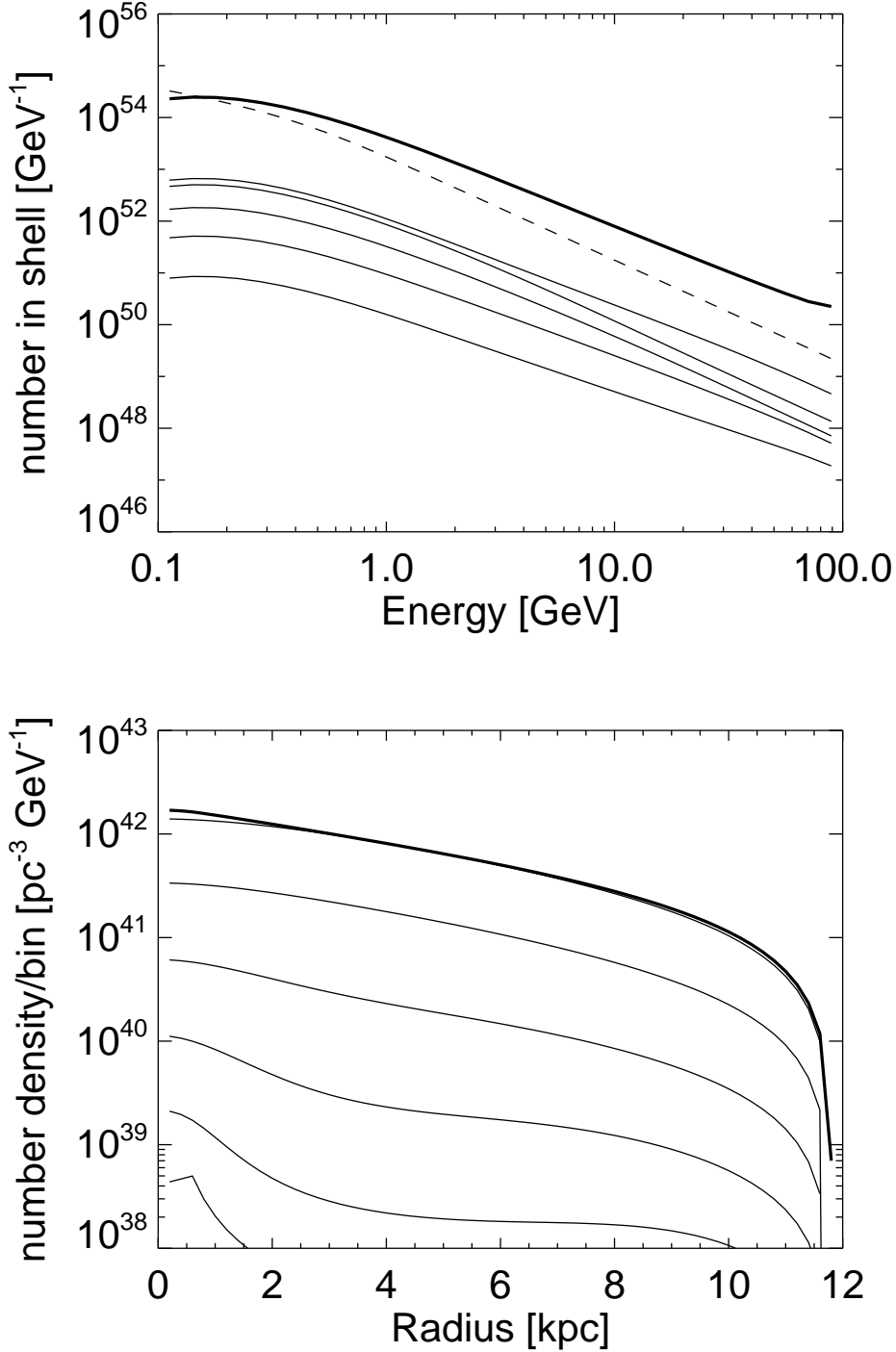}
\figcaption{(\emph{top}) Model 2 electron energy distribution for
all electrons (\emph{thick line}) and spherical shells as in
Fig. \ref{fig_simple}
from the numerical calculation in
\S\ref{sec_numerical}.  For comparison, the analytic solution for
Model 1 is evaluated for all electrons (\emph{dashed
 line}) as before.  The slopes are somewhat steeper at high energy,
but could be made flatter by changing the assumptions about the
boundaries of the diffusion zone.
(\emph{bottom}) Number density ($\pc^{-3}$) as a function of
Galacto-centric radius for all electrons (\emph{thick line}) and for
bins of energy 0.35, 1.1, 3.5, 11, 35, and 89 GeV (\emph{thin
 lines, top to bottom}). 
The highest energy bin is truncated at 600 pc.  In both panels the
figures are normalized assuming a total annihilation rate of $4\times
10^{38}\s^{-1}$. 
\label{fig_density}
}
\end{figure}

\begin{figure}[tb]
\epsscale{0.8}
\dfplot{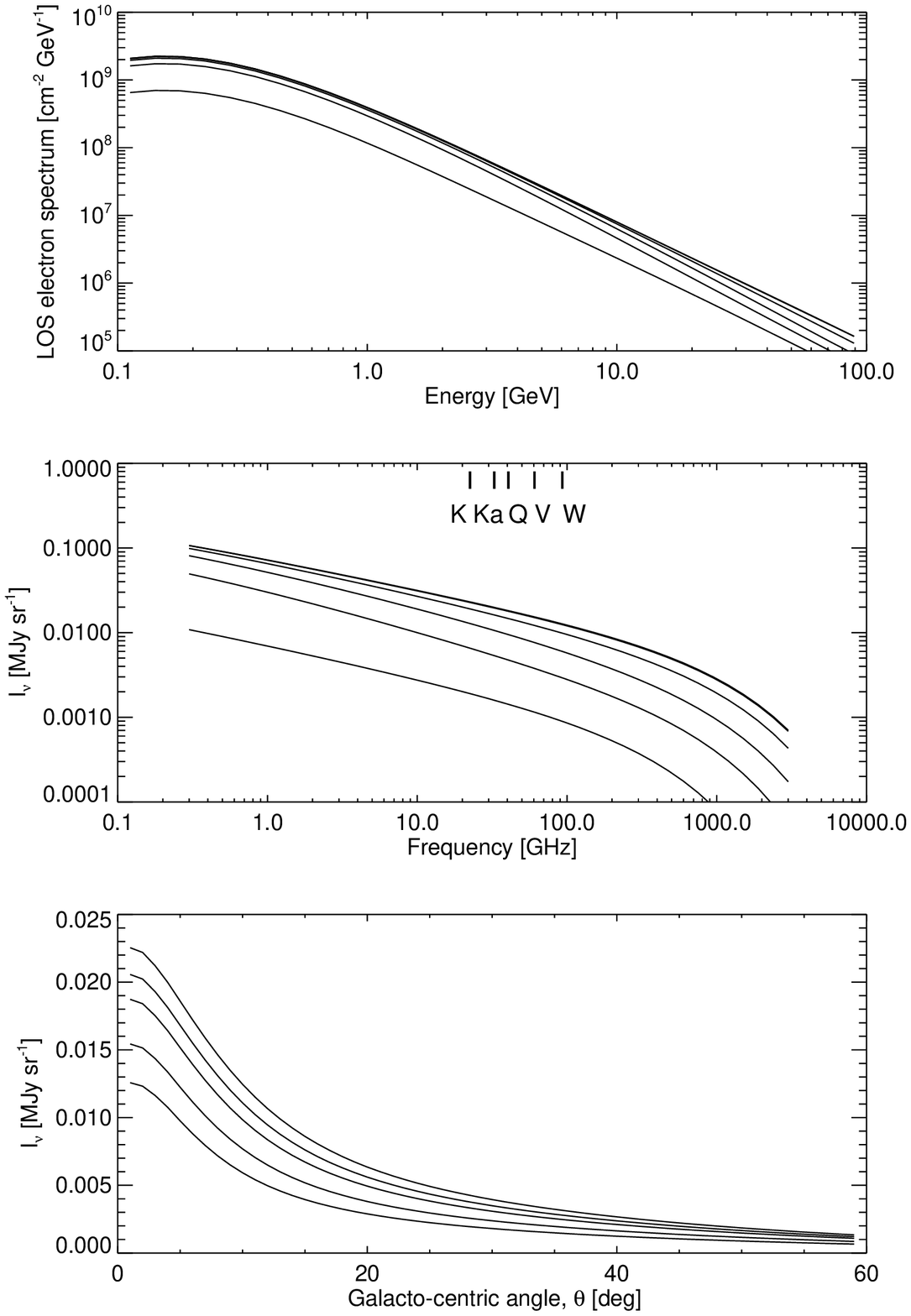}
\figcaption{
(\emph{a}) Line of sight electron energy distribution
[$\cm^{-2}\GeV^{-1}$].  Curves are shown for Galacto-centric angle
$\theta = (2,5,10,20,$ and $50\degree)$.  The numerical solution rolls
off at lower
energy because of electron escape from the diffusion zone. 
(\emph{b}) Specific Intensity vs. frequency for lines of sight at
the same angles.  The spectra nearest the Galactic
center are flattest, as discussed in \S\ref{sec_analytic}. 
(\emph{c}) Specific Intensity vs. $\theta$ for frequency bins centered
at 23.8, 30.0, 37.8, 59.9, and 94.9 GHz, close to the \WMAP\
frequencies, (\emph{highest to lowest}).
These calculations assume a total annihilation rate of $4\times
10^{38}\s^{-1}$, and that all power goes into $e^+e^-$.  These plots
must be rescaled by a factor of 0.25 to obtain the model shown in
Figure \ref{fig_dm_haze}.
\label{fig_synch}
}
\end{figure}

\subsection{Electron Energy Distribution}
\label{sec_result}
The electron energy distribution for constant $A$ and $K$ (model 1) is
shown in Figure \ref{fig_simple}, both as a function of energy,
plotted for different radii, and as a function of radius, plotted for
different energies.  Results for the analytic solution
(\S\ref{sec_analytic}) are also shown, to demonstrate that diffusion
from the edges of the numerical grid is unimportant for energies $E\ga
1\GeV$.  Because these higher energies are the ones relevant for
synchrotron emission at $\nu > 23\GHz$, the numerical solution is
adequate for a comparison to \WMAP. 

The more realistic Model 2, with variable $K$ and $A$, yields a
somewhat steeper electron energy distribution (Fig. \ref{fig_density}), but
results are qualitatively similar.  The electron spectrum is
integrated along each line of sight, and the synchrotron spectrum
computed (Figure \ref{fig_synch}).  The corresponding synchrotron
spectrum for Model 1 is very similar.  The power available from
annihilation (a total annihilation rate of $4\times 10^{38}\s^{-1}$
for the Galaxy; see Appendix A) is the same in both models, so it is
no surprise that the total power output in synchrotron is similar. 

\begin{figure}[tb]
\epsscale{1.0}
\dfplot{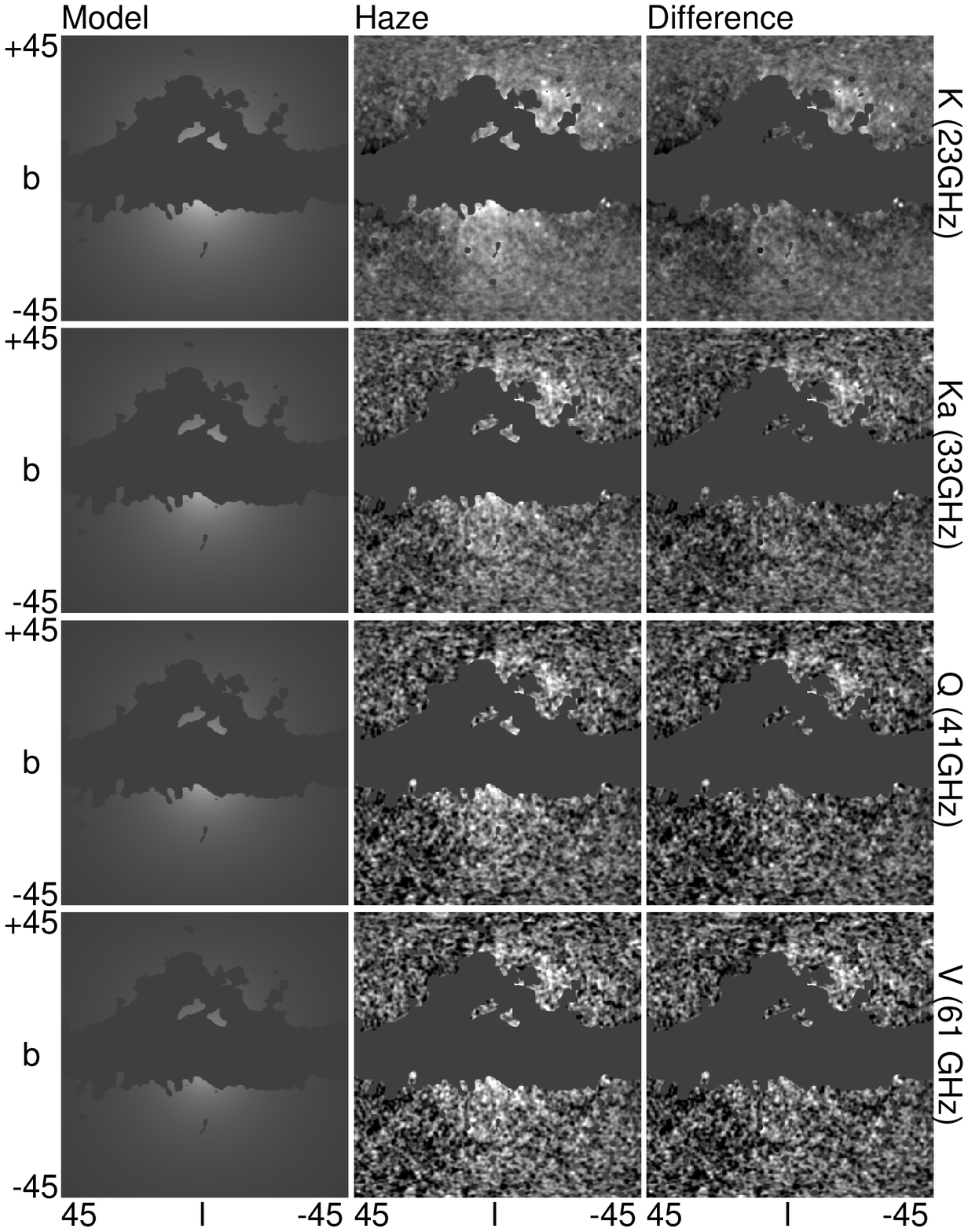}
\figcaption{WMAP haze within $45\degree$ of the Galactic center.
Columns contain the emission from the model described in
\S\ref{sec_numerical} (\emph{left}), haze from Finkbeiner (2004)
(\emph{center}), and haze minus model (\emph{right}).  Each row
corresponds to a
\WMAP\ band, with W-band absent due to lower signal-to-noise ratio.
Pixels with known point sources, high extinction (more than 1 mag at
\Halpha), or bright \Halpha ($> 25$ R) are masked.  The color table in
each image goes from $-2$
(\emph{black}) to 6 (\emph{white}) kJy/sr.  One kJy/sr is $67\microK$
at $23\GHz$. 
\label{fig_dm_haze}
}
\end{figure}

The radial profile of synchrotron emission is shown in 
Figure \ref{fig_synch}c for the \WMAP\ bands,
but these assume that all annihilation power appears as
ultra-relativistic $e^+e^-$.  Comparison of the model with the
observed \WMAP\ haze residual (Finkbeiner 2004) in
Figure \ref{fig_dm_haze} indicates that a factor of 0.25
is appropriate, i.e.  25\% of the annihilation power appears as
synchrotron emission.  For a different neutralino mass, the annihilation
power varies as $m_\chi^{-1}$ so this result would appear to put a
hard upper limit of a few hundred GeV on the mass.  For $m_\chi <
100\GeV$ a smaller fraction of the power would be required, but most
models with $m_\chi \la 80\GeV$ violate accelerator constraints.  

Of course, halo clumpiness of the kind invoked by de Boer \etal\
(2004) can increase the available power, and allow greater $m_\chi$
and/or more diffusion losses off the edge of the diffusion zone.

\subsection{X-rays and $\gamma$-rays}

The hard $\gamma$ rays produced by inverse Compton scattering of
starlight photons have an energy density similar to the microwave
synchrotron emission (\S\ref{sec_photons}).  The softer $\gamma$ rays
from upscattered CMB photons are expected to be subdominant in the
inner galaxy, but perhaps still detectable. 
Extensive searches for these signal have been carried out, and
detections so far are suggestive.

The EGRET data from the \emph{Compton Gamma Ray Observatory} at $0.1 <
E_\gamma < 10\GeV$ have long been claimed to show an excess of
emission in the inner Galaxy over that expected from cosmic ray
interaction with the ISM (Hunter \etal\ 1997; Strong, Moskalenko, \&
Reimer 2000).  The hard $\gamma$ excess has remained enigmatic, though
Strong \etal\ (2000) were able to fit the observed spectrum by
assuming a much flatter electron energy distribution above 1 GeV than
that observed in the cosmic ray population near Earth.  Recently, de
Boer \etal\ (2004) have considered dark matter annihilation as a
source for these hot electrons and found reasonable agreement with
$\gamma$ ray data and other constraints.  Some of the de Boer \etal\
assumptions about dark matter structure in the Galaxy may be
controversial (a high degree of clumpiness is assumed), but it is
interesting that both groups conclude that a very hot
electron spectrum is present.  Note that the absence of clumpiness
would imply a lower $\gamma$ ray production, and would not violate the
observations; there could be other sources as well. 

An excess at lower energies observed by COMPTEL is at roughly the
energy expected for scattered CMB photons; however the observed amplitude is
higher by one or two orders of magnitude.  The INTEGRAL $\gamma$ ray
observatory has revealed numerous soft $\gamma$ point sources in the
inner galaxy that could account for this emission (Lebrun \etal\
2004).  The fact that these sources are resolved leaves open the
possibility that the upscattered CMB may be detected at a lower level
with more sensitive instruments in the future.
A detection of CMB scattering would be much easier to interpret than
the putative starlight scattering, since the CMB photon density is
isotropic and known.
Starlight scattering calculations are significantly complicated by
the fact that both energy and direction distributions must be known.

A great deal of work remains to tie the information provided
by the observation of the \WMAP\ microwave haze to the important
constraints already imposed by the $\gamma$ ray data.

\section{SUMMARY}
We have attempted to constrain models of self-annihilating dark
matter by searching for synchrotron emission from the annihilation
products, assumed to be ultra-relativistic $e^+e^-$ pairs.
We refer to the dark matter particle as the
supersymmetric neutralino, and the mass as $m_\chi$, even though the
results are of generic interest for any stable dark matter particle 
that annihilates into ultra-relativistic $e^+e^-$.

The Galactic emission found by Finkbeiner (2004) in the \WMAP\
data in excess of the expected foreground Galactic ISM signal may be a
signature of such dark matter annihilation.  It has the spectrum,
morphology, and total power expected for the simple model given above,
which uses reasonable values for the parameters of the dark matter
halo density, particle mass, and cross section; parameters that agree
with other cosmological and accelerator constraints.  Additional tests
of this model are possible, given the X-rays and $\gamma$-rays
produced by inverse-Compton scattered CMB and starlight photons,
respectively.  Current data are consistent with the model, but provides
only loose constraints.

In addition to better X-ray and $\gamma$-ray data, sensitive microwave
observations of other edge-on spiral galaxies (e.g. NGC 891 or NGC 4565) might
reveal a similar signal, and verify these results.  Tests of Milky Way
satellite dwarf galaxies have also been suggested (Baltz \& Wai 2004)
but the signal from these may be too weak for current technology. 

While the dark matter annihilation interpretation is exciting, it must
be emphasized that other explanations are possible.  A hard
electron energy spectrum could be produced by novel physics
around the supermassive black hole at the Galactic center.  Although
observations of other galaxies have not revealed such a flat spectrum
of emission, they may have lacked the necessary resolution to 
separate the various emission components.

A more thorough investigation is required before declaring that the
energetic population of electrons result from dark matter
annihilation, and indeed, astrophysical observations alone are
unlikely to make an entirely convincing case unless an annihilation
line or sharp energy cut-off is found in $\gamma$ rays.  Complementary
data from the
LHC (Ellis \etal\ 2004) and other particle accelerators will play an
essential role in making the case for a new particle.  The vastly
improved data expected over the next few years, both experimental and
observational, may finally provide an observational window into one of
the great mysteries of modern cosmology, dark matter.

I am indebted to Jim Gunn and Bruce Draine for help and encouragement;
David Schlegel, Robert Lupton, and Nikhil Padmanabhan for inspiration,
technical advice, and helpful skepticism; Carl Heiles for advice on
the Galactic B-field and other matters; Marc Davis for enlightening
discussions; and Amber Miller for reminding me what an important problem
dark matter annihilation is.  Gus and Flora Schultz's kind hospitality
allowed me to work out the essentials of this problem while visiting their
home in Berkeley. 
This research made use of the NASA Astrophysics Data
System (ADS) and the IDL Astronomy User's Library at
Goddard\footnote{http://idlastro.gsfc.nasa.gov/}.
I am supported by NASA LTSA grant NAG5-12972, the Russell Fellowship,
and the Cotsen Fellowship of the Society of Fellows in the Liberal
Arts at Princeton University. 

\clearpage

\appendix
\section{TOTAL MILKY WAY ANNIHILATION}
We assume the DM density in the Galactic halo obeys the NFW
density profile, 
\BE
\rho_{NFW}(r) = \frac{\rho_0r_s^3}{r(r^2 + r_s^2)}
\EE
with a core radius $r_s=20\kpc$ and
$\rho_0=5.6\times10^{-25}\g\cm^{-3}$, 
making the density in the Solar neighborhood 
$\rho(R_\sun) = 6.5\times10^{-25}\g\cm^{-3}$ and
$M_{gal} = 2\times10^{12} M_\sun$.
The annihilation rate per volume is
\BE
\Gamma(r) = \left(\frac{\rho_{NFW}(r)}{m_\chi}\right)^2 \sigmav
\EE
For Majorana particles, the annihilation rate is $\sigmav
n_\chi^2/2$, but in each annihilation, two particles are removed,
which cancels the factor of 2 in the annihilation rate. 

Note that the integral of $\Gamma(r)$ over all space (total number of
annihilations per time) is
\BE
4\pi\int_0^{\infty} \Gamma(r) r^2 dr =
\pi^2\left(\rho_0/m_\chi\right)^2r_s^3 \sigmav
\EE
which for the assumed parameters of $m_\chi=100\GeV$, $\sigmav =
2\times 10^{-26} \cm^3 \s^{-1}$, $\rho_0=5.6\times10^{-25}\g\cm^{-3}$,
and $r_s=20$kpc, is $4.6\times10^{38}\s^{-1}$ for the whole Galaxy.
Assuming that the $e^+e^-$ pair creation rate is proportional to
$\Gamma$, we define the efficiency $\eta$ as the number of
relativistic $e^-$ (or $e^+$; hereafter simply called electrons)
created per neutralino annihilation, so
that the creation rate is $\eta\Gamma(r)$.  The electron density is
obtained by convolving this density function with the Green function
obtained in Eq (\ref{eq_green}).  In general, such a convolution looks like
\BE
h(E,r)=\int d\phi~d\theta r'^2 \sin\theta dr' \Gamma(r') G(|\vecx'-\vecx|)
\EE
and since both functions depend only on radius, we can assume $r$ to
be on the $z$ axis such that the angle between $r$ and $r'$ is
$\theta$, 
then 
\BE
n(E,r)=\int d\phi~d\theta r'^2 \sin\theta dr' \Gamma(r') G(\sqrt{r^2+r'^2-2rr'\cos\theta})
\EE
Inserting the functions gives
\BE
n(E,r)=\frac{\rho_0^2r_s^6\sigmav}{AE^2m_\chi^2(2\pi\sigma^2)^{3/2}}
\int d\phi~d\theta r'^2 \sin\theta dr'
\frac{1}{r'^2(r'^2 + r_s^2)^2}
\exp(-(r^2+r'^2-2rr'\cos\theta)/2\sigma^2)
\EE
where $\sigma_A$ is the neutralino self-annihilation cross section,
not to be confused with the Gaussian width $\sigma$.
Integrating over $\theta$ and $\phi$ we get
\BE
n(E,r)=\frac{\rho_0^2r_s^6\sigmav}{AE^2m_\chi^2(2\pi\sigma^2)^{1/2}}
\int dr'
\frac{1}{r r'(r'^2 + r_s^2)^2} \left[
\exp\left(-\frac{(r-r')^2)}{2\sigma^2}\right) -
\exp\left(-\frac{(r+r')^2)}{2\sigma^2}\right) \right]
\EE
This quantity is number density per energy. 
Integrating $n(E,r)|dE/dt|$ over the whole Galaxy, we obtain a total
annihilation rate of $4\times10^{38}\s^{-1}$

\end{document}